\documentclass[twocolumn,twoside,fleqn,showpacs,showkeys,pra,aps,10pt,superscriptaddress,final]{article}

\usepackage{hep}
\graphicspath{{figs/}}      

\def\refname{References}

\def\journalname{??}

\def\@pacs@name{PACS numbers: }%
\def\@keys@name{Keywords: }%

\def\Dated@name{Dated: }%
\def\Received@name{Received }%
\def\Revised@name{Revised }%
\def\Accepted@name{Accepted }%
\def\Published@name{Published }%
\def\address{\replace@command\address\affiliation}%
\def\altaddress{\replace@command\altaddress\altaffiliation}%

\definecolor{orangec}{cmyk}{.24,.91,.96,.18}
\definecolor{orangecc}{cmyk}{.24,.94,.96,.18}
\definecolor{oorangec}{cmyk}{.8,.2,.5,.4}
\definecolor{ooorangec}{cmyk}{1,.9,0.08,.04}      

\definecolor{orangec}{cmyk}{.15,.7,.96,.0}
\definecolor{orangecc}{cmyk}{.15,.7,.96,.0}

\newcommand{\ODDBOX}{\noindent\raisebox{-13.3cm}{\includegraphics{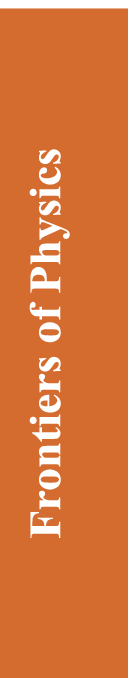}}}
\newcommand{\EVENBOX}{\noindent\raisebox{-13.3cm}{\includegraphics{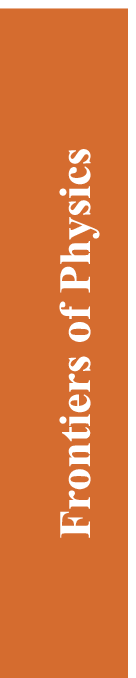}}}

\newfont{\yihao}{cmb10 at 18pt}
\newcommand{\Yihao}{\fontsize{18pt}{13.5pt}\selectfont}
\newcommand{\xiaosi}{\fontsize{11.5pt}{13.5pt}\selectfont}         

\newfont{\xbt}{cmb10 at 12pt}

\def\frontmatter@title@format{%
    \centering%
    \usefont{T1}{fradmcn}{m}{n}\yihao}%
\def\@keys@name{{\color{ooorangec}\bf Keywords~~}}%
\def\@pacs@name{{\color{ooorangec}\bf PACS numbers~~}\vspace{2mm}}%
\def\frontmatter@authorformat{\vspace{5mm}\centering\bf}%

\newcommand{\catchline}[2]{
	{\vspace*{-16.4mm}\small%
	\noindent #1\\%
	\noindent #2\\[-2mm]%
    {\color{orangec}{\rule{\textwidth}{.5pt}}}\\[2mm]
    {\color{orangec}{\Yihao\textbf{\textsc{\Papertype}}}}\\[6mm]
    }\relax\par
}

\newcommand{\doi}[1]{DOI {#1}}
\renewcommand{\title}[1]
{\vspace*{-5mm}\begin{center}
{\Yihao\bf #1}
\end{center}
}

\renewcommand{\author}[1]
{\vspace*{0mm}
\begin{center}
{\bf #1}
\end{center}
}
\newcommand{\add}[1]{\begin{center}{\small\it #1}\end{center}}

\newcommand{\abs}[1]{
\begin{center}
\parbox[t]{156mm}{\noindent\color{oorangec}#1}
\end{center}}

\newcommand{\keywords}[1]{
\begin{center}
\parbox[t]{156mm}{\noindent{\bf\color{ooorangec}Keywords}\ \ #1}
\end{center}}

\newcommand{\pacsnumbers}[1]{
\begin{center}
\parbox[t]{156mm}{\noindent{\bf\color{ooorangec}PACS numbers}\ \ #1\vspace*{5mm}}
\end{center}}

\newcommand{\acknowledgements}[1]{\vspace*{4mm}\noindent{\renewcommand{\baselinestretch}{1.05}\footnotesize{\color{ooorangec}\bf Acknowledgements}\quad{#1}}}

\def\journalname{??}
\def\volumenumber#1{\gdef\@volumenumber{#1}}%
\def\@volumenumber{}%
\def\issuenumber#1{\gdef\@issuenumber{#1}}%
\def\@issuenumber{}%
\def\volumeyear#1{\gdef\@volumeyear{#1}}%
\def\@volumeyear{}%

\renewcommand\thesection{\arabic{section}}
\renewcommand\thesubsection{\arabic{section}.\arabic{subsection}}
\renewcommand\thesubsubsection{\arabic{section}.\arabic{subsection}.\arabic{subsubsection}}

\titleformat{\section}[hang]{\color{ooorangec}\vspace*{-1.2mm}\titlerule\vspace{1mm}\large\usefont{T1}{fradmcn}{m}{n}\xbt}{\thesection}{1em}{}
\titlespacing{\section}{0mm}{8mm}{5mm}

\titleformat{\subsection}{\normalfont\normalsize\color{ooorangec}}{\thesubsection}{1em}{}
\titlespacing{\subsection}{0mm}{5mm}{3mm}

\titleformat{\subsubsection}{\normalfont\normalsize\it\color{ooorangec}}{\thesubsubsection}{1em}{}
\titlespacing{\subsubsection}{0mm}{3mm}{3mm}

\titlecontents{section}
[0em] {\normalfont} {\thecontentslabel\quad} {\thecontentslabel}
{\titlerule*[.7pc]{}\contentspage}

\titlecontents{subsection}
[1.5em] {\normalfont} {\thecontentslabel\quad} {\thecontentslabel}
{\titlerule*[.7pc]{}\contentspage}

\titlecontents{subsubsection}
[4em] {\normalfont} {\thecontentslabel\quad} {\thecontentslabel}
{\titlerule*[.7pc]{}\contentspage}

\usepackage[small]{caption2}

\newlength{\halfpagewidth}
\divide\halfpagewidth by 1

\usepackage{hyperref}
\usepackage{mathrsfs}
\usepackage{ulem}

\begin{document}

	\newcommand{\Papertype}{\sc Research article} 
	\def\volumeyear{2020} 
	\def\volumenumber{???}  
	\def\issuenumber{??}
	\def\journalname{Front. Phys.} 
	\newcommand{\doiurl}{???} 
	\newcommand{\allauthors}{Fen Lyu, Yan -Zhi Meng, Zhen-Fan Tang. et al.}
	\twocolumn[
	\begin{@twocolumnfalse}
		\catchline{\journalname~\volumenumber,~\issuenumber~(\volumeyear)}{\doi{\doiurl}} 
		\thispagestyle{firstpage}
		

		\title{A comparison between repeating bursts of FRB 121102 and giant pulses from Crab pulsar and its applications}
		
		\author{Fen Lyu$^{1,2*}$, Yan-Zhi Meng$^{1,3}$, Zhen-Fan Tang$^{1}$, Ye Li$^{1,4}$, Jun-Jie Wei$^{1}$, Jin-Jun Geng$^{3,5**}$, Lin Lin$^{6}$, Can-Min Deng$^{7***}$, Xue-Feng Wu$^{1,8****}$}
		
		\add{1. Purple Mountain Observatory, Chinese Academy of Sciences, Nanjing 210023, China\\
			2. University of Chinese Academy of Sciences, Beijing 100049, China\\
			3. School of Astronomy and Space Science, Nanjing University, Nanjing 210023, Jiangsu, China \\
			4. Kavli Institute for Astronomy and Astrophysics, Peking University, Beijing 100871, China \\
			5. Institute of Astronomy and Astrophysics, University of T\"ubingen, Auf der Morgenstelle 10, D-72076,T\"ubingen\\
			6. Department of Astronomy, Beijing Normal University, Beijing 100875, China\\
			7.  Department of Astronomy, University of Science and Technology of China, Hefei 230026, Anhui, China\\
			8. School of Astronomy and Space Sciences, University of Science and Technology of China, Hefei 230026, Anhui, China\\
			Corresponding author \ E-mail: $^{*}$lyufen@pmo.ac.cn, $^{**}$gengjinjun@nju.edu.cn,$^{***}$dengcm@ustc.edu.cn,$^{****}$ xfwu@pmo.ac.cn\\}

\abs{There are some similarities between bursts of repeating fast radio bursts (FRBs) and giant pulses (GPs) of pulsars. To explore possible relations between them, we study the cumulative energy distributions of these two phenomena using the observations of repeating FRB 121102 and the GPs of Crab pulsar. We find that the power-law slope of GPs (with fluence $\geq$130 Jy ms) is  $2.85\pm0.10$. The energy distribution of FRB 121102 can be well fitted by a smooth broken power-law function. For the bursts of FRB 121102 above the break energy (1.22 $\times 10^{37}$ erg), the best-fitting slope is  $2.90_{-0.44}^{+0.55}$, similar to the index of GPs at the same observing frequency ($\sim$1.4 GHz). We further discuss the physical origin of the repeating FRB 121102 in the framework of the super GPs model. And we find that the super GPs model involving a millisecond pulsar is workable and favored for explaining FRB 121102 despite that the magnetar burst model is more popular.}


\pacsnumbers{97.60.Gb, 98.70.Dk, 02.70.Rr}

\vspace*{-6mm}

\end{@twocolumnfalse}
]

\section{
	Introduction} \label{sec:intro}

Fast Radio Bursts (FRBs) are mysterious, bright astronomical millisecond-duration radio pulses \cite{2007Sci...318..777L,2018MNRAS.478.1209F,2018Natur.553..182M}, occasionally discovered in pulsar searches. Up to date, more than one hundred FRBs have been reported
\cite{2017MNRAS.469.4465P}\footnote{https://www.frbcat.org}. They are expected to be of cosmological origin due to high dispersion measures (DM, $10^{2}-10^{3}$ pc cm$^{-3}$) in excess of the DM contribution from the Milky Way. And it is further confirmed by the localization of the host galaxies
\cite{2017Natur.541...58C,2019Sci...365..565B,2019Sci...366..231P,2020Natur.577..190M}.

Among the moderately observed large sample of FRBs, most of them are one-off events, and only twenty of them sporadically show repeating bursts \cite{2016Natur.531..202S,2019Natur.566..235C,2019Natur.566..230C}. Moreover, FRB 180916.J0158+65 detected by CHIME \cite{2020Natur.582..351C} exhibits a $\sim$16 day period acitvity with unknown mechanism. Most recently, a bright FRB 200428 has been reported to be spatially coincident with the galactic Soft Gamma-ray Repeater (SGR) 1935+2154 \cite{2020arXiv200510324T,2020arXiv200511071L,2020ApJ...898L..29M,2020arXiv200511479L}, also associated with a hard X-ray burst \cite{2020arXiv200511071L,2020arXiv200512164T,2020ApJ...898L..29M,2020arXiv200511178R}.
The repetition of FRB 200428 is very rare, which may be different from other extragalactic repeating FRBs \cite{2020arXiv200510324T}. 

The progenitors of FRBs are still under heated debates, and many theoretical models have been proposed (see \cite{2019PhR...821....1P} and references therein\footnote{https://frbtheorycat.org/index.php/Main\_Page}), despite some of them have confronted serious challenges. Repeating FRBs are the best candidates to explore physical nature due to its repetition. FRB 121102 is the first observed repeating FRB and has been detected in the frequency range from 600 MHz \cite{2019ApJ...882L..18J} to 8 GHz \cite{2018ApJ...863....2G}. For repeating FRBs, some models involving catastrophic processes have been ruled out, e.g., binary neutron star (NS) mergers \cite{2012ApJ...755...80P,2016ApJ...822L...7W}, and binary white dwarf mergers \cite{2013ApJ...776L..39K}. Many non-catastrophic models have been put forward, such as the flaring magnetars model \cite{2010vaoa.conf..129P,2014MNRAS.442L...9L,
	2017ApJ...843L..26B,
	2019MNRAS.485.4091M}, accretion process by a neutron star in a binary system 
\cite{2020MNRAS.497.1543G}, the interaction of NS with an asteroid belt
\cite{2016ApJ...829...27D}, the interaction between the NS magnetosphere and cosmic winds \cite{2017ApJ...836L..32Z}, and the giant pulses (GPs) from pulsars \cite{2016MNRAS.457..232C,2016MNRAS.458L..19C}. The central engine of FRB 121102 is questionable, which can be constrained from the persistent radio nebula of FRB 121102 for the magnetar-type model \cite{2017ApJ...839L..20C,
		2017ApJ...842...34W,
		2018MNRAS.481.2407M,
		2019ApJ...885..149Y,
		2020ApJ...892..135W}. 

GPs were first discovered in the Crab pulsar (PSR B0531+21), and then were found in several energetic pulsars (see Table 1 of  \cite{2019SCPMA..6279511W} and reference therein). 
Different from the regular pulses, the typical duration of GPs is very short, with no clear periodicity, ranging from nanoseconds to microseconds. The flux density could be 2 or 4 orders of magnitude higher than the average pulse-integrated flux of regular energetic pulses (e.g., \cite{1995ApJ...453..433L,2012A&A...538A...7K}). Among the pulsars emitting the GPs, the largest value of the magnetic field at the light cylinder $B_{\rm lc}$ are the Crab pulsar and PSR B1937+21 \cite{2019SCPMA..6279511W}. The Crab pulsar is a remarkable source emitting numerous GPs and simultaneously has been detected across a broadband range of radio frequencies from 20 MHz \cite{2013ApJ...768..136E} to 46 GHz \cite{2015ApJ...802..130H}. The distinctive feature of GPs is the energy of GPs obeys a power-law distribution, whereas the regular ones follow a Gaussian or log-normal distribution (e.g.,\cite{1995ApJ...453..433L,2004ApJ...612..375C}), indicating that the emission mechanism of GPs may be different from that of regular pulses. Meanwhile, GPs are also crucial to understand the radiation mechanism of pulsars. 

Repeating FRBs and GPs from pulsars share similarities in many observational aspects, such as
brightness temperature, duration time, sub-bursts, polarization, and frequency drift. 
Moreover, the emission mechanism of both of them is coherent radiation, supporting that they may have a similar physical origin. Actually, in some theoretical models, it has been suggested that FRBs can be treated as GPs from young rapidly rotating pulsars \cite{2016MNRAS.457..232C,2016MNRAS.458L..19C}.


The statistical properties, such as the energy distributions, of repeating FRBs and GPs might help us to understand FRBs progenitors. In this paper, we focus on the statistical properties of the repeating FRB 121102 and GPs from the Crab pulsar.
%
The paper is organized as follows. In Section 2, we introduce the sample selection. 
The statistical analysis of the isotropic-equvivalent energy $E$ is shown in Section 3. Discussion and conclusions are given in Section 4 and Section 5. We assume a flat $\Lambda$CDM universe with $\Omega_{m}=0.27$, and $H_{0}=70\,{\rm km}\,{\rm s}^{-1}{\rm Mpc}^{-1}$ throughout this work.

\section{
	Sample selection}\label{sec:Sample}
The Crab Nebula pulsar is well known for its GPs, with detected GPs numerous enough for detailed studies (e.g., \cite{1995ApJ...453..433L,2012A&A...538A...7K,2019MNRAS.490L..12B}). Here, GPs refer to GPs generated by the main pulse (MP) and interpulse (IP) phase.

In our analysis, we adopt a complete sample of 1153 bright GPs (with fluence $\geq$130 Jy ms to avoid the incomplete effect near the detection threshold) from the Crab pulsar \cite{2019MNRAS.490L..12B}. These GPs were detected by the 15-m telescope in Pune, India (National Centre for Radio Astrophysics) in a 260-hours observation from February to April 2019. The observation was conducted between 1280 MHz and 1380 MHz, with 65 MHz bandwidth usable. Each GP was detected with a signal-to-noise ratio (S/N) of $\geq10$.

Among 20 repeating FRBs have been reported, only two of repeating FRBs have host galaxy identified and redshift measured (FRB 121102 and FRB 180916.J0158+65; \cite{2017Natur.541...58C,2020Natur.582..351C}). FRB 121102 is the first and currently the best representative of repeating bursts. More than one hundred bursts from FRB 121102 have been detected by several different surveys, including Green Bank Telescope (GBT) in C-band (4-8 GHz; \cite{2018ApJ...866..149Z}) and Arecibo telescope at 1.4 GHz \cite{2019ApJ...877L..19G}. The repeating FRB 121102 has a very large and variable Faraday rotation measure ($\mathrm{RM} \sim 10^{5}\, \mathrm{rad}  \,\mathrm{m}^{-2}$) which means the burst inhabits in an extreme magneto-ionic environment \cite{2018Natur.553..182M}.


To compare with the Crab GPs sample under the same condition, we use the sample of FRB 121102 detected at almost the same frequency \cite{2019ApJ...877L..19G}. The observing frequency of Arecibo in the source frame is $1.4 \ast(1+{z})$ GHz. Due to the redshift 0.193 for FRB 121102 \cite{2017ApJ...834L...7T}, the frequency in the source rest frame approximatively is $\sim1.4$ GHz. In this sample, there are 41 bursts from FRB 121102 observed by the 305-m Arecibo telescope. The data is from two observations taken on 2016 September 13/09:47:07 and 14/09:50:12, and lasting 5967 s and 5545 s. Each observation detected 18 and 23 bursts with S/N $\ge$10, respectively. Moreover, we also investigate the energy distribution of the repeating FRB 121102 detected by GBT telescope at C-band \cite{2018ApJ...866..149Z}. There are 93 bursts in total in 5 hours with a convolutional neural network technique.

\section{
	The cumulative energy $E$ distributions}\label{sec:analyses}

Energy is a crucial parameter to identify the progenitor and to explore the radiation mechanism. We perform an analysis on the energy distribution of the repeating FRB 121102 and GPs to explore the progenitor of FRB 121102, despite there are many works devoted to the statistical properties of the repeating FRBs \cite{2016MNRAS.461L.122L, 2017JCAP...03..023W,2018MNRAS.475.5109O,2018ApJ...866..149Z,2019ApJ...883...40L}.

The isotropic-equvivalent energy $E$ in the source rest frame within the observing bandwidth for the selected sample can be calculated by
\begin{equation}
E=4\pi d_{L}^{2}f_{\nu} \Delta \nu / (1+z)\;, \label{Eiso}
\end{equation}
where $d_{L}$ is the luminosity distance of the source, with 2.0 kpc for the Crab pulsar, 
$f_{\nu}$ is the fluence density of each burst, and $\Delta \nu $ is the corresponding
bandwidth of every burst. In this paper, the bandwidth for FRB 121102 is derived
from the burst frequency edges ($f_{\rm high}$ and $f_{\rm low}$) from the table
given in \cite{2019ApJ...877L..19G}.

Here we adopt a simple power-law model, $N(>E)\propto E^{-\alpha}$ to fit the cumulative energy distributions of bright GPs (fluence $\geq$ 130 Jy ms) from the Crab pulsar. 
While for the repeating FRB 121102, we fit it by the smooth broken power-law function, i.e.,
\begin{equation}	
N_{\mathrm{cum}}(>E)=A\, \left[\left(\frac{E}{E_{\mathrm{b}}}\right)^{\alpha_{1} \omega}+\left(\frac{E}{E_{\mathrm{b}}}\right)^{\alpha_{2} \omega}\right]^{-1 / \omega}\label{bpl},
\end{equation}
where $A$ is the ampltitude, $\alpha_{1}, \alpha_{2}$ are slopes of the two segments, $E_{\mathrm{b}}$ is the break energy, $\omega$ is the sharpness (or smoothness) of the break of energy distribution, and it usually can be set as a constant ($\omega$ = 3).

We can now obtain the goodness-of-fit $\chi_{\mathrm{d}. \mathrm{o.f}}$ for the difference between the observed cumulative distribution $N_{\mathrm{cum}, \mathrm{obs}}$ and the theoretical distribution function $N_{\mathrm{cum}}$ for the cumulative distribution (with $n_{\mathrm{par}}$ = 4 parameters, $E_{0}$, $\alpha_{1}$, $E_{\mathrm{b}}$, $\alpha_{2}$ )
\begin{small}
\begin{equation}
\chi_{\mathrm{d.o.f}}=\sqrt{\frac{1}{\left(N_{\mathrm{tot}}-n_{\mathrm{par}}\right)} \sum_{i=1}^{N_{\mathrm{tot}}}\frac{\left[N_{\mathrm{cum}}\left(E_{i}\right)-N_{\mathrm{cum}, \mathrm{obs}}\left(E_{i}\right)\right]^{2}}{\sigma_{\mathrm{cum}, i}^{2}}},
\end{equation}
\end{small}
where $N_{\mathrm{tot}}$ is the number of bursts dectected by the same instrument at the same frequency for the same source. Note that the uncertainty of $N_{\mathrm{cum}}(>E)$ is taken as $\sigma_{\mathrm{cum}, i}$ = $\sqrt{N_{\mathrm{cum}, \mathrm{obs}}\left(E_{i}\right)}$. The best-fitting parameters and the uncertainties can be obtained by minimizing $\chi_{\mathrm{d.o.f}}$ via the python package emcee (a Bayesian MCMC method by sampling the affine-invariant for Markov Chain Monte Carlo chains, \cite{2013PASP..125..306F}). 

When fitting variables via a MCMC approach, we assume the various parameters in the fitting function are uniform distribution in the prior probability distribution. Here are the range of parameters for every sample: 1 < $\alpha $< 5 for GPs; 50 < $A$ < 200, -2.5 <$\alpha_{1}$ < -0.5, 36 < lg$E_\mathrm{b}$ <38 and 2 < $\alpha_{2}$  < 5 for Arecibo sample; 50 < $A$< 200, -3< $\alpha_{1}$ < -0.01, 36 < lg$E_\mathrm{b}$ <38.5 and 1< $\alpha_{2}$ < 3 for GBT sample.

\begin{figure}
	\centering
	\includegraphics[width=0.85\linewidth]{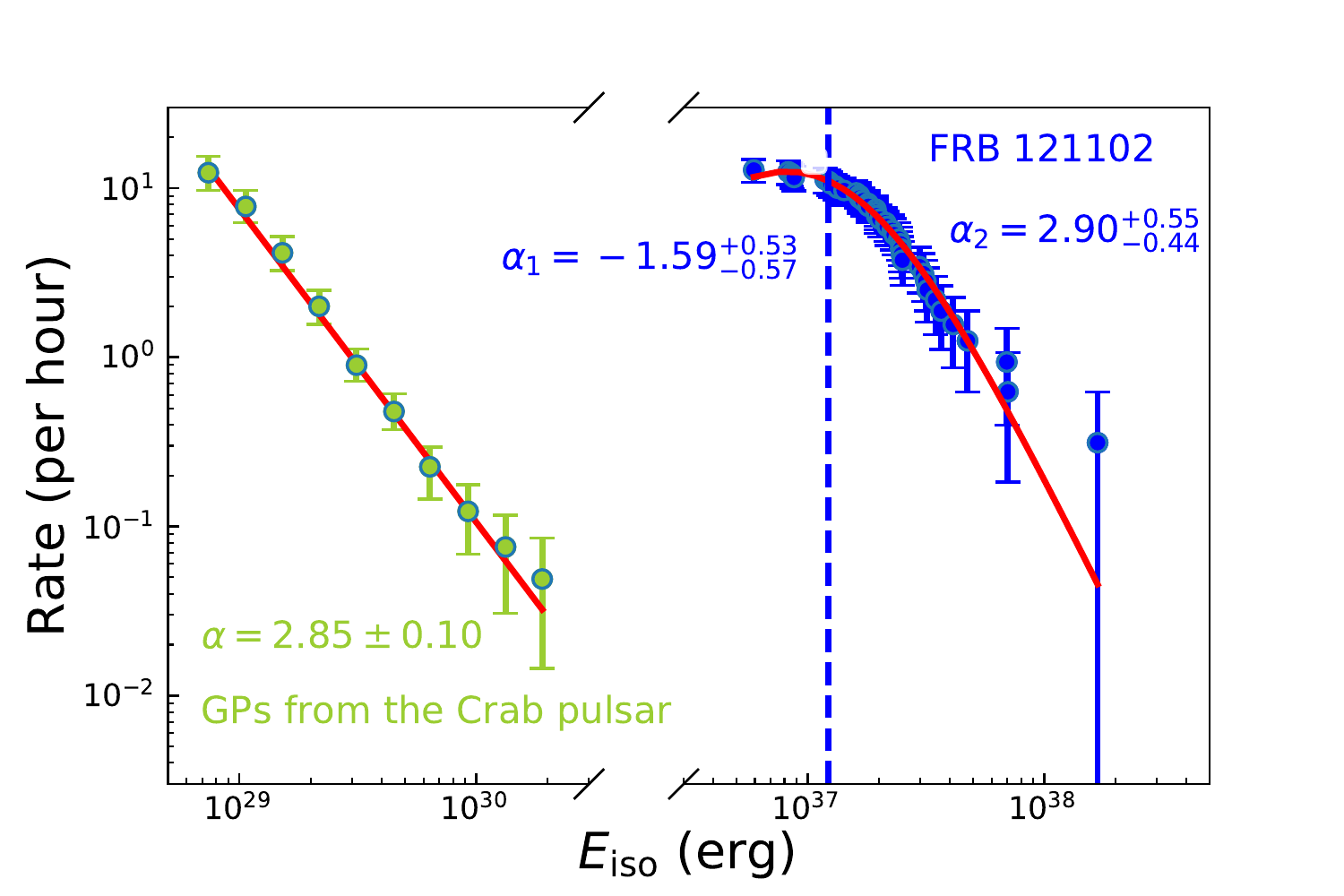}
	\includegraphics[width=0.818\linewidth]{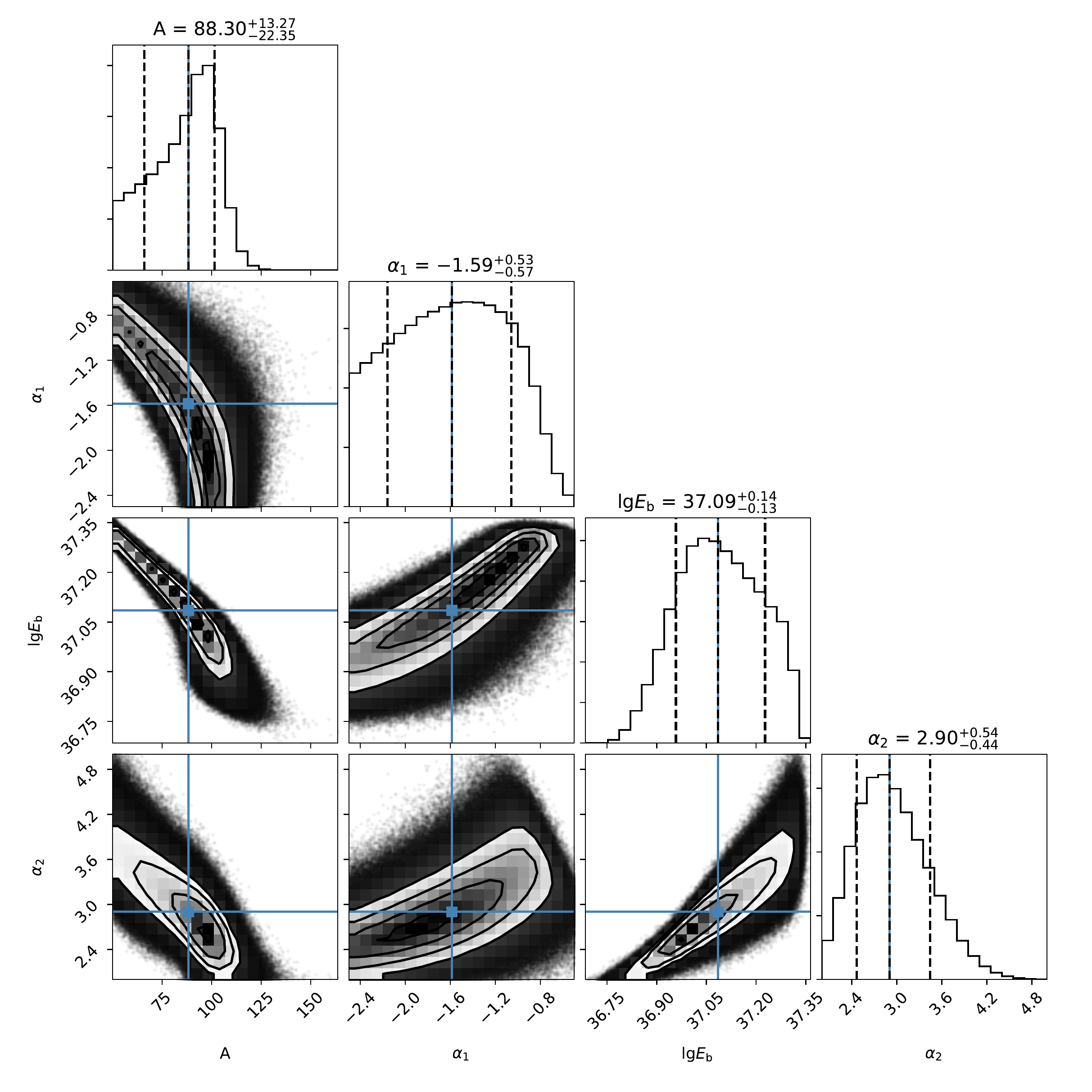}
	\caption{Upper panel: The cumulative energy $E_{\rm iso}$ distributions of the GPs from the Crab pulsar (yellow-green), FRB 121102 detected by Arecibo (blue circle). The vertical blue dotted line represents the break energy of FRB 121102 fitted, and the red line represents the best fitting line; Lower panel: 1D, 2D posterior marginalized probability distributions of the fitting parameters obtained using  MCMC method for FRB 121102 detected by Arecibo telescope. The contours in the 2D plots are 68$\%$ and 95$\%$ confidence intervals from inside to outside.}
	\label{Fig_1}
\end{figure}

GPs are generally considered as the power-law tail of the regular pulses for the pulsar. In the upper panel of Figure \ref{Fig_1}, the yellow-green points represent the GPs from the Crab pulsar, which is found to be well fitted by the power-law model. The rollover does not appear at the high-energy end, while it does show for the flux density distribution as stated in some previous studies \cite{1995ApJ...453..433L,2004ApJ...612..375C}. The best-fit slope is $ 2.85\pm0.10$ ($\chi_{\mathrm{d.o.f}}$ = 0.45), which is well consistent with that of the fluence distribution ($2.8 \pm 0.3$) reported in \cite{2019MNRAS.490L..12B}. In Figure\ref{Fig_1}, we present the cumulative energy ($E$) distribution for 41 Arecibo bursts from FRB 121102 (blue spots). Fitted by the broken power-law function given in equation (\ref{bpl}), the best fitting results for indices of the lower and higher energy segment are $-1.59_{-0.57}^{+0.53}$ and $2.90_{-0.44}^{+0.55}$ ($\chi_{\mathrm{d.o.f}}$ = 0.36), respectively, while the break energy $E_{b}$ is $1.22_{-0.31}^{+0.46}$ $\times 10^{37}$ erg (blue dashed line). In \cite{2019ApJ...877L..19G}, the threshold energy was set at $E_{\rm th}=2\times 10^{37}$ erg considering the sensitivity limit and the observed turnover. Above the $E_{\rm th}$,  the best fit is $\frac{d R}{d E} \propto E^{-2.8}$, which is generally consistent with our fitting result within the uncertainty. All the uncertainties quote a 68\% confidence interval for the best-fitted parameters.

\begin{figure}
	\centering
	\includegraphics[width=0.82\linewidth]{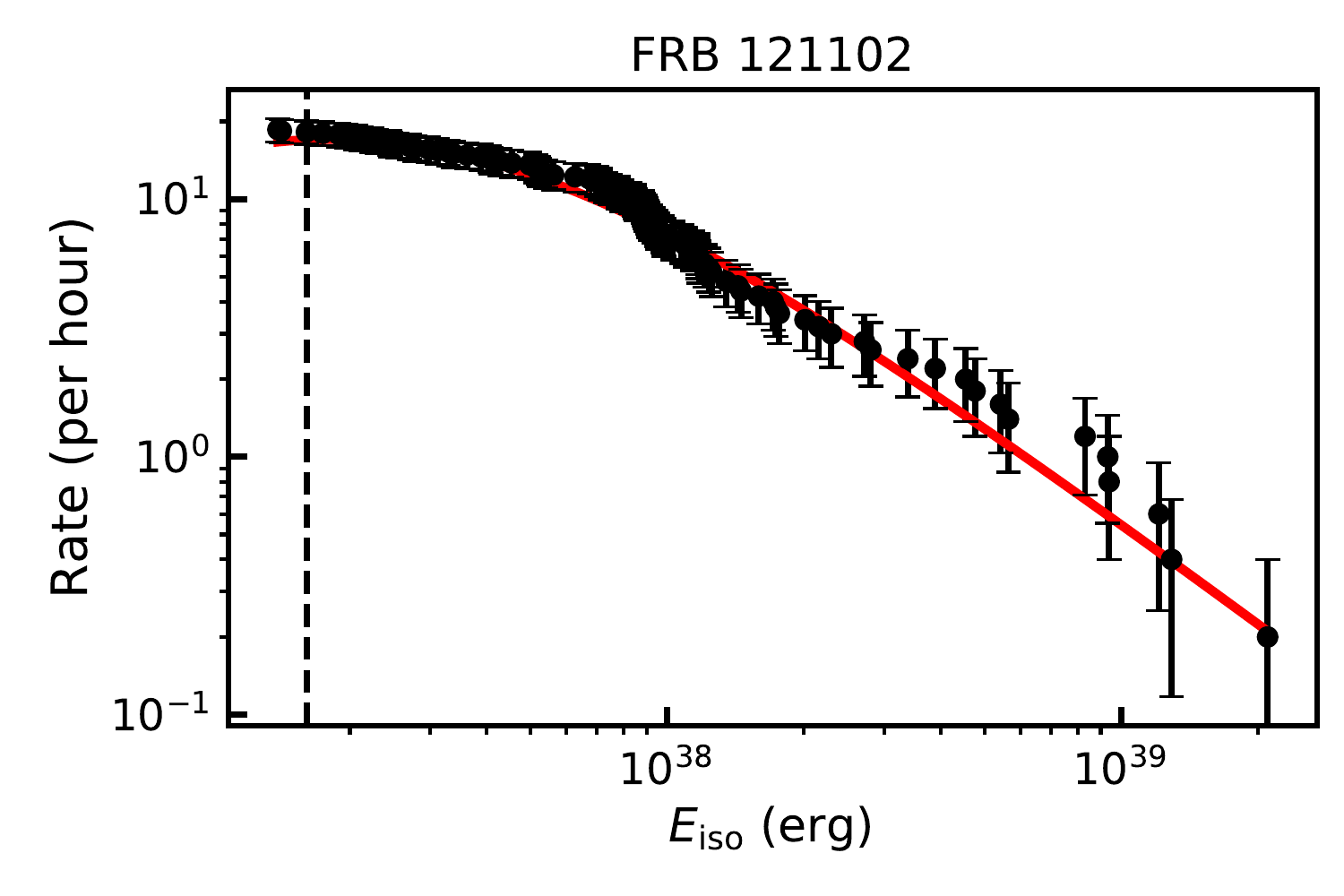}
	\includegraphics[width=0.82\linewidth]{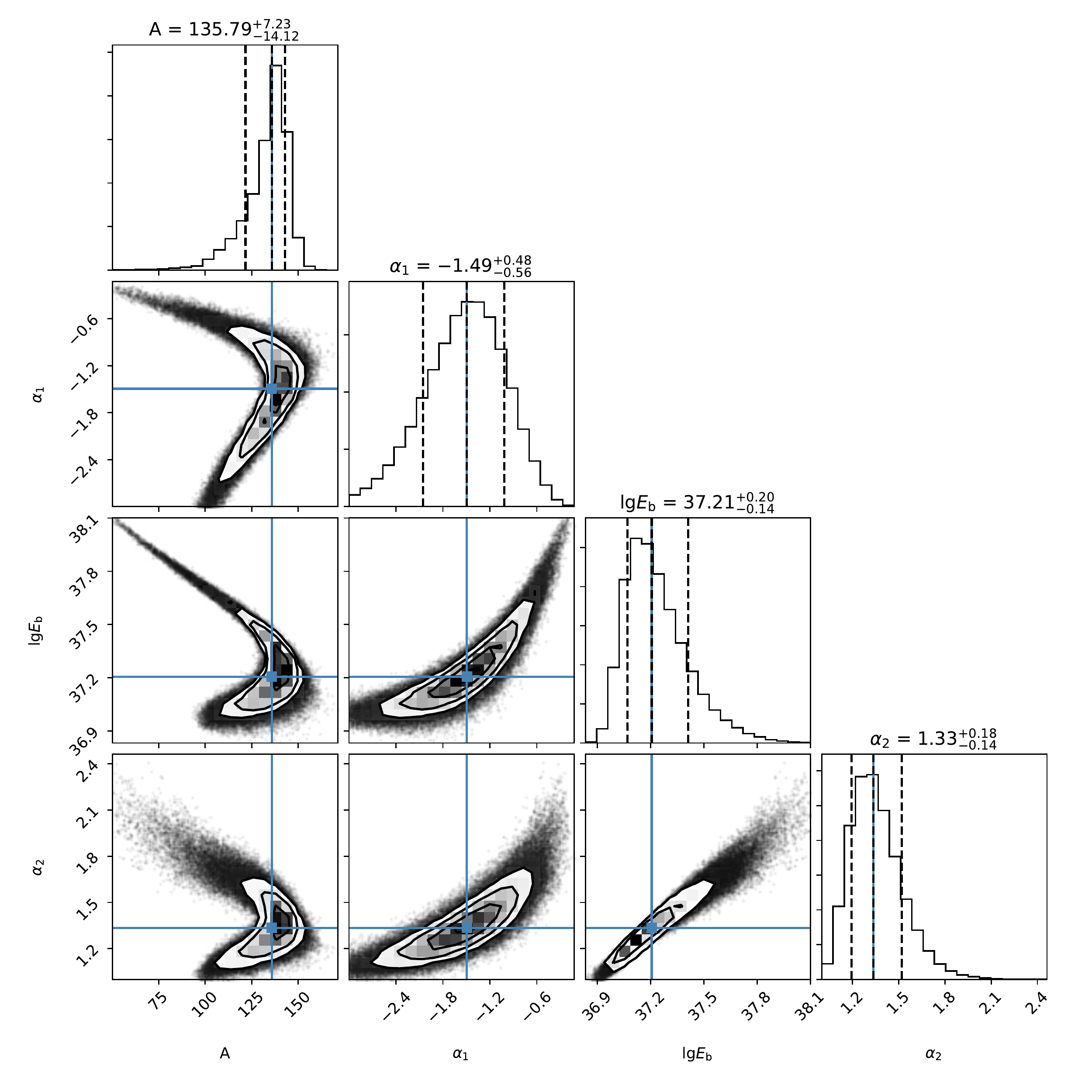}
	\caption{Upper panel: The cumulative energy distribution of FRB 121102 detected by GBT telescope at C-band, the solid red line denotes the best fit line using equation (\ref{bpl}) with the best-fit parameters labeled in the lower panel, and the black dashed line is the break energy; Lower panel: 1D, 2D posterior marginalized probability distributions of the fitting parameters obtained using  MCMC method. The contours in the 2D plots are 68$\%$ and 95$\%$ confidence intervals from inside to outside.}
	\label{fig:gbt1}
\end{figure}

To investigate what causes the different power indices of the energy distribution of FRB 121102 detected by the Arecibo telescope at L-band and that by GBT telescope at C- band and the meaning of the break energy $E_{b}$ in Figure \ref{Fig_1}, we also fit the energy distribution for 93 bursts from FRB 121102 by GBT telescope with broken power-law function. As shown in Figure \ref{fig:gbt1}, the best fits for the lower and higher energy are $-1.49_{-0.56}^{+0.48}$ and $1.33_{-0.14}^{+0.18}$ ($\chi_{\mathrm{d.o.f}}$ = 0.63), respectively, while the break energy is $1.60_{-0.44}^{+0.94}$ $\times 10^{37}$ erg. 

Interestingly, with the same observed frequency ($\sim $ 1.4 GHz), FRB 121102 and Crab GPs share the power-law form of similar indices ($\sim$ 2.8) (Figure \ref{Fig_1}). At a different observed frequency (C-band, 4-8 GHz), the energy distribution of FRB 121102 detected by the GBT telescope is flatter (Figure \ref{fig:gbt1}) than that of the Arecibo telescope with more energetic bursts.

There is a lack of bursts below near the break energy $\sim 1.60 \times 10^{37}$ erg (the black dashed line) due to the instrument sensitivity and observational limit. So maybe the incomplete sampling below the threshold is the major reason that evokes the different power-law indices of energy distribution for FRB 121102 in Figure\ref{Fig_1} and Figure \ref{fig:gbt1}. The power-law index of energy distribution with repeating FRB 121102 may depend on the energy band, the burst is more energetic at a higher observational frequency (see Figure \ref{Fig_1} and Figure \ref{fig:gbt1}),  which can be further tested by detecting more bursts at different energy bands. Note that the energy distribution of GPs might also depend on the energy band \cite{1995ApJ...453..433L,2004ApJ...612..375C}, and can have the index $\sim $ 1.8 for some pulsars (such as PSR 0631+1036; \cite{2008ApJ...672.1103M}).

The power-law index of energy distribution and the statistical characteristics of the other observable quantities of FRB121102 are claimed to be similar to those of soft gamma repeater bursts from magnetars \cite{2020MNRAS.491.1498C}. In addition, there are several statistical analyses \cite{2017JCAP...03..023W,2017ChPhC..41f5104C,2017MNRAS.471.2517W,2018ApJ...852..140W,2019arXiv190311895Z,2020MNRAS.491.1498C,2020FrPhy..1614501L} dedicated to the self-organizing critical (SOC) study, especially some researches \cite{2017JCAP...03..023W,2019arXiv190311895Z,2020MNRAS.491.1498C,2020MNRAS.491.2156L} that indicate that the repeating FRB 121102 is a SOC system \cite{1987PhRvL..59..381B}. Thus, the repeating FRB 121102, is likely to be an SOC system due to some kind of instability driven above the instability threshold value. If so, the break energy in Figure \ref{Fig_1} should be the threshold energy, which determines the trigger mechanism of the system. It is an important quantity that affects the slope of the power-law in energy distribution, which is also discussed in \cite{2019arXiv190311895Z}.


Interestingly, the power-law slope of FRB 121102 omitting bursts that fall bellow the break energy is generally consistent with that of the GPs from the Crab as shown in Figure \ref{Fig_1}, which implies they may share a similar origin. If FRB 121102 is extragalactic GPs, which energies can power it? We will discuss this in the next section below. 

\section{Discussion}
\label{sec:discuss}
As mentioned in the introduction, neutron stars are the most popular FRB central engine. Especially the super GPs model involving pulsars 
\cite{2016MNRAS.457..232C,
	2016MNRAS.458L..19C,
	2016MNRAS.462..941L} and the giant flare model involving magnetars \cite{2013arXiv1307.4924P,
	2014MNRAS.442L...9L,
	2017ApJ...843L..26B,
	2019MNRAS.485.4091M,
	2020ApJ...896..142B,
	2020MNRAS.494.4627M} have been widely discussed. Specifically for FRB 121102, it was found that its host is a dwarf galaxy with a high star formation rate \cite{2017Natur.541...58C,2017ApJ...834L...7T}, which implies that the central engine of FRB 121102 may locate in the remnant of the death of a massive star. Moreover, the polarization observations give that the Faraday rotation measure of FRB 121102 is surprisingly high $\sim 10^5 ~\rm{rad ~ m^{-2}}$, indicating that FRB 121102 should be surrounded by a strongly magnetized environment \cite{2018Natur.553..182M}. These observational evidences support that the central engine of FRB 121102 may be a magnetized NS. In the following, we explore the physical origin of the repeating FRB 121102 in the framework of the super GPs model.

As shown in the upper panel of figure \ref{Fig_1}, the typical energy of the FRB 121102 and GPs are $10^{37} \sim 10^{38}$ erg and $10^{29} \sim 10^{30}$ erg. On the other hand, the typical time scale of the FRB 121102 and GPs are ms and $ \mu$s, hence the typical luminosity are $10^{40} \sim 10^{41}$ erg/s and $10^{35} \sim 10^{36}$ erg/s, respectively.
Following \cite{2017ApJ...838L..13L}, based on the typical luminosity of FRB 121102 and that of GPs, we have 
\begin{equation}
\xi = \frac{{L}_{\mathrm{FRB}}}{{L}_{\mathrm{GP}}} \sim   10^{5} .
\end{equation}
Since the GPs are powered by the spin down of the pulsar  $L_{\rm{sd}}$, 
the constraint on the magnetic field strength  $B_{\mathrm{FRB}}$ of the FRB source is
\begin{equation}
B \approx  0.3~ \xi_5^{1/2} B_{\rm{Crab}} \left(\frac{P_{\rm{-3}}}{P_{\rm{Crab}}}\right)^{2},
\end{equation}
By scaling the power of the Crab GPs to the level of FRB 121102 i.e. $L_{\rm{sd,FRB}}=\frac{{L}_{\mathrm{FRB}}}{{L}_{\mathrm{GP}}} L_{\rm{sd,GP}}$,
where $B_{\rm{Crab}}\sim 4 \times 10^{12}$ G and $P_{\rm{Crab}} \sim 33$ ms are the magnetic field strength and the period of the Crab pulsar, respectively \cite{2000ApJ...535..365K}.
It is seen that the millisecond pulsars may produce GPs with luminosity comparable to FRB 121102. On the other hand, the active time scale is the spin-down time scale of the pulsar, that is
\begin{equation}
t_{\rm active} \sim 50 ~ \xi_5^{-1}I_{45}  R_{6}^{-6}  B_{\rm{Crab}}^{-2} P_{\rm{Crab}}^{4}P_{-3}^{-2}~ \mathrm{yr}.
\end{equation}
which can in principle explain FRB 121102 so far. Considering the condition of $t_{\mathrm{active}}$ > 7 yrs, we constraints that the period of the initial spin $P$ should be smaller than 3 ms. In addition, if the radiation efficiency of FRB $\eta_{\rm{FRB}}$ is higher than that of GPs $\eta_{\rm{GP}}$, for example an order of magnitude higher, then we would have a longer active time scale $t_{\rm active} \sim 500~ (\eta_{\rm{FRB}}/10\eta_{\rm{GP}})$. Then the constraint to the period of the initial spin is $P \leq 8 $ ms in this case. For comparison, the constraint from the persistent radio nebula gives $P \leq 7~ \mathrm{ms}$ \cite{2019ApJ...885..149Y}.

Moreover, we note that both repeating FRBs and GPs have high brightness temperature ($\geq10^{35}$ K), inferring that their radiation mechanism should be coherent emission \cite{2018ApJ...868...31Y,2020ApJ...901L..13Y}. GPs from Crab pulsar and FRBs 121102 share similar complex pulses morphology and a similar repetition rate in a single pulsar ($%
\sim $ 1-100 per hour, see Figure~\ref{Fig_1}). Furthermore, similar to FRBs \cite{2015MNRAS.447..246P,2015Natur.528..523M}, GPs have strong polarization signals, which is either left-handed or right-handed polarization \cite{2007whsn.conf...68S} between linear polarization and circular polarization. Besides, spectral structures in repeating FRBs resemble those seen in Crab GPs \cite{2007ApJ...670..693H}. Therefore, the similarities between the GPs and FRB 121102 possibly link to the same origin and mechanism of these two phenomena.

\section{Conclusions}\label{sec:Conclude}
In this paper, we have made a comparison of the energy distributions between the repeating FRBs and the GPs from the Crab pulsar. We find that FRB 121102 and Crab GPs share similar power-law indices in the energy distribution at the same observed frequency $\sim $1.4 GHz\footnote{Although the fitting results of GBT data and Arecibo data for repeating FRB 121102 are not consistent with each other, we have discussed some possible reasons for the inconsistency in section 3 such as under different observation frequency, a lack of the bursts below the break energy for Arecibo data, or the intrinsic physical mechanism.}.
Furthermore, we explored the physical origin of the repeating FRB 121102 in the framework of the giant pulses model. We find that the millisecond pulsar is possible to produce super GPs with luminosity comparable to FRB 121102 in a reasonable active time scale. Therefore, we argue that the super GPs model is workable and favored to explain FRB 121102 despite that the magnetar burst model is more popular.

\acknowledgements{
	We thank the anonymous referee for constructive comments and suggestions that improved the paper. This work is partially supported by the National Natural Science Foundation of China (grant Nos. 11673068, 11725314, U1831122, 11903019, 11533003, and 11703002), the Youth Innovation Promotion Association (2017366), the Key Research Program of Frontier Sciences (grant Nos. QYZDB-SSW-SYS005 and ZDBS-LY-7014), the Strategic Priority Research Program “Multi-waveband gravitational wave universe” (grant No. XDB23000000) of the Chinese Academy of Sciences, the China Post- doctoral Science Foundation (Nos. 2018M631242 and 2020M671876), the Fundamental Research Funds for the Central Universities, and the National Postdoctoral Program for Innovative Talents (grant No. BX20200164).
}


\raggedend

\begin{small}

\end{small}


\end{document}